# Student Conceptions of Holography

# Schülervorstellungen zur Holographie

Martin E. Horn, Helmut F. Mikelskis

University of Potsdam, Physics Education Research Group,
Am Neuen Palais 10, 14469 Potsdam, Germany
E-Mail: marhorn@rz.uni-potsdam.de – mikelskis@rz.uni-potsdam.de

**Abstract**

In a lesson on holography a tension is to be found between the ray model, the wave model, the phasor model and the particle model. Lessons depend on the previous experiences of the students, the intentions of the teacher as well as on other parameters and therefore provide a diversity in emphasis while explaining holographic effects.

As class discussion concerning selected areas of interference optics and specifically holography develops in a questioning manner and with limited intervention from the teacher, numerous student conceptions with regard to interference optics and different model conceptions of light can be identified. Student interviews in which these discovered concepts are explored in greater detail can provide in depth information. Primary student conceptions, which were gathered in the main research conducted for a dissertation project will be presented here and analysed with a view to their didactic consequences.

**Kurzfassung**

Ein Unterricht zur Holographie liegt im Spannungsfeld von Strahlenmodell, Wellenmodell, Zeigermodell und Teilchenmodell. In Abhängigkeit von Vorerfahrungen der Schülerinnen und Schüler sowie Lehrerintentionen und weiteren Parametern liefert er unterschiedlich akzentuierte Erklärungsmuster holographischer Effekte.

In einem fragend-entwickelnden Unterrichtsgespräch zu ausgewählten Gebieten der Interferenzoptik und speziell der Holographie mit nur schwacher Intervention seitens des Lehrers lassen sich zahlreiche Schülervorstellungen zur Interferenzoptik und zu unterschiedlichen Modellvorstellungen zum Licht identifizieren. Vertiefte Informationen liefern Schülerinterviews, in denen die aufgefundenen Konzepte hinterfragt werden. Wesentliche Schülervorstellungen, die im Rahmen der Hauptuntersuchung eines Dissertationsvorhabens erhoben wurden, werden präsentiert und hinsichtlich ihrer didaktischen Konsequenzen analysiert.

## Contents

## Inhalt

## 1. Holography at school

Even though holographic images can be discovered more frequently in the everyday and holography is a topic oriented toward application for upper level optics, holography is only rarely included in physics lessons. An analysis of school books reveals that one of the reasons for this reserve on the part of the teachers can be the unsystematic and didactically difficult representation of holography in school books [8]. Moreover, the topic is considered complex by teachers and the didactic reassessment of holography, which meanwhile spans a spectrum from curriculum recommendations for physics lessons at the intermediate level [4] to a clarification of quantum physical processes in view of holography at the secondary school level II [3], has not yet reached schools.

Nonetheless, holography is a topic that is particularly motivating to students and provides an opportunity to constructively work out and use model representations of light during lessons. Within this sphere of tensions between the models (i.e. ray model, wave model, phasor model as well as the particle model of light) student conceptions will develop and will be superimposed with preconcepts, which are elementary for comprehending the learning process that is underway.

Thus a lesson unit of 12 school hours was developed at the University of Potsdam [10] in order to examine model building processes in interference optics. The following basic structure forms the foundation of this lesson unit:

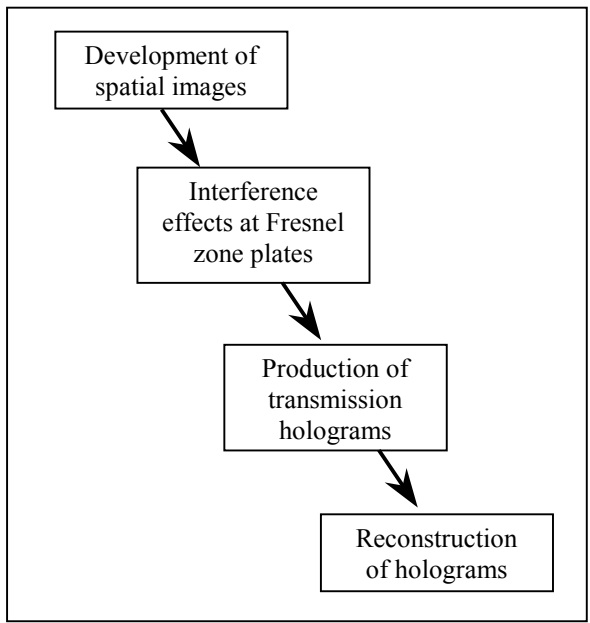

***Figure 1:*** *Structure of the lesson unit on holography.*

One of the focal points of this proposal is the implementation of simulation programs in ray optics [7] and interference optics.

## 1. Holographie in der Schule

Obwohl holographische Bilder immer häufiger im Alltag anzutreffen sind und die Holographie ein anwendungsorientiertes Thema für die Oberstufenoptik darstellt, findet eine Behandlung der Holographie im Physikunterricht derzeit nur selten statt. Wie eine Schulbuchanalyse zeigt, kann eine der Ursachen dieser Zurückhaltung der Lehrerinnen und Lehrer in der unsystematischen und didaktisch teilweise problematischen Darstellung der Holographie in Schulbüchern liegen [8]. Auch gilt das Thema als komplex, und die didaktische Aufarbeitung der Holographie, die mittlerweile eine Spannweite zwischen Unterrichtsvorschlägen für den Physikunterricht der Mittelstufe [4] bis hin zu Erläuterungen quantenphysikalischer Prozesse am Beispiel der Holographie in der Sekundarstufe II [3] umfasst, hat die schulische Unterrichtsgestaltung noch nicht erreicht.

Dennoch ist die Holographie ein Schülerinnen und Schüler hochgradig motivierendes Thema, das sich dazu anbietet, gerade auch die Modellvorstellungen des Lichts im Unterricht konstruktiv erarbeiten und anwenden zu lassen. Im Spannungsfeld dieser Modelle (Strahlenmodell, Wellenmodell, Zeigermodell und Teilchenmodell des Lichts) werden sich Schülervorstellungen bilden und mit Präkonzepten überlagern, die für ein Verständnis der ablaufenden Lernprozesse elementar sind.

An der Universität Potsdam wurde deshalb eine Unterrichtseinheit im Umfang von 12 Schulstunden entwickelt [10], um Modellbildungsprozesse im Bereich der Interferenzoptik zu untersuchen. Der Gestaltung dieser Unterrichtseinheit liegt die in Abbildung 1 dargestellte Grobstruktur zugrunde.

Ein Schwerpunkt dieses Unterrichtsvorschlags ist der Einsatz von Simulationsprogrammen im Bereich der Strahlenoptik [7] und der Interferenzoptik.

## 2. Zielsetzung und Durchführung der Untersuchung

Da in Deutschland der Optikunterricht der Sekundarstufe II Interferenzerscheinungen zumeist im Kontext des Wellenmodells behandelt, während in der Sekundarstufe I eine Vorprägung der Schülerinnen und Schüler durch das Strahlenmodell erfolgt, sind kognitive Konflikte nicht vermeidbar. Zur Untersuchung dieser Prozesse wurden und werden in Potsdam neben der Entwicklung, Erprobung und Evaluation des Unterrichtskonzepts hinsichtlich Lernzuwachs, Motivation der Lernenden und Praktikabilität der Unterrichtsdurchführung sowie den Auswirkungen des Einsatzes der Computer-Simulationsprogramme unter anderem die folgenden wissenschaftlichen Fragestellungen bearbeitet:

- Welche Rolle spielen Präkonzepte und Alltagsvorstellungen bei einem Unterricht zur Holographie?

## 2. Objectives and implementation of research

Cognitive conflicts are unavoidable since lessons on optics at the secondary school level II in Germany cover interference primarily within the context of the wave model while at the secondary school level I the students are predisposed to the ray model. In addition to development, testing and evaluation of lesson concepts with regard to advancement in learning, motivation of the students, practicability of lesson implementation as well as the impact of using computer simulation programs, the following scientific questions have been and are being currently considered in Potsdam for the purpose of researching these processes. These questions are:

- What are the roles of preconceptions and everyday perceptions in teaching holography?
- What is the impact of models used by the students on the comprehension of the basic subject matter of holography and what degree of distinction exists between alternative model conceptions?

The developed lesson concept was implemented in the 2000/2001 school year in the general and extension courses of the secondary school level II in Brandenburg schools and videotaped. The students of these classes also prepared concept maps and participated in a pre- and a final test. The questions on the tests were related the motivation and attitude of the students, their previous knowledge of optics, specifically interference optics, as well as their understanding of models. After the lesson a part of the students was questioned in more depth about aspects of their model conceptions.

The main feature of this paper will be the evaluation of these interviews as well as primary sections of lessons; a statistical and empirical analysis of the questionnaires will follow. Moreover, it is planned to conduct another control test at a later point in time in order to examine the long-term effects on learning.

## 3. Student conceptions at secondary school level I

In a preliminary examination [9] at a Berlin comprehensive school, conceptions concerning the term "hologram" were gathered from 166 students between grades 8 and 10. The results showed that approximately two-thirds of the pupils did not know anything about a "hologram."

Allegedly slightly more than ten percent of the participating students did reply that they had seen a real hologram before. These students were able to name the primary characteristics of three-dimensionality of holographic effects using their emergent technical terminology. The third part however that consisted of more than 20 percent of the students was able to deliver a considerably detailed description. With the

- Wie wirken sich die von den Lernenden verwendeten Modelle auf das Verstehen grundlegender physikalischer Sachverhalte aus, und welche Trennschärfe ist zwischen alternativen Modellvorstellungen vorhanden?

Das entwickelte Unterrichtskonzept wurde im Schuljahr 2000/2001 in Grund- und Leistungskursen der Sekundarstufe II an Brandenburger Gymnasien unterrichtet und videographiert. Ebenso fertigten die

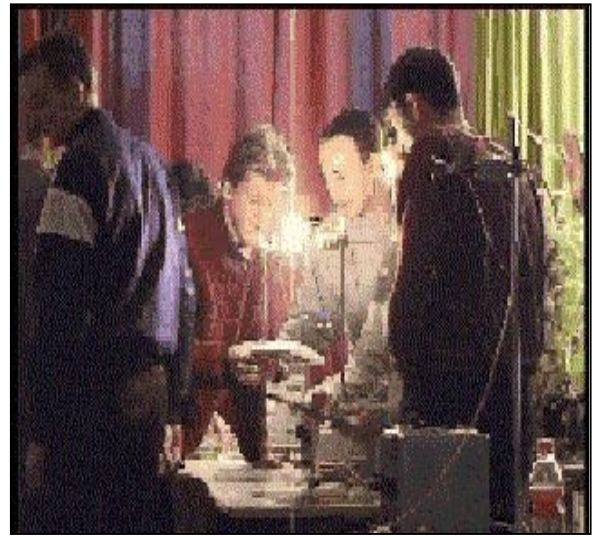

***Figure 2:*** *Students observe spatial effects on carved figures during a lesson on holography.*

Schülerinnen und Schüler dieser Kurse Concept Maps an und nahmen jeweils an einem Vor- und einem Nachtest teil, wobei sich die Testfragen auf Motivation und Einstellung der Schüler, auf das vorhandene Wissen im Bereich der Optik und insbesondere der Interferenzoptik sowie auf das Modellverständnis der Befragten bezogen. Nach Abschluss des Unterrichts wurde ein Teil der Schüler in Interviews zu Aspekten ihrer Modellvorstellungen vertieft befragt.

Die Auswertung dieser Interviews sowie zentrale Unterrichtspassagen bilden die Grundlage dieses Beitrags, wobei eine statistisch-empirische Analyse der Fragebögen noch folgen wird. Ebenso ist mit einem Teil der Schülerpopulation die Durchführung eines späteren Kontrolltests zur Ermittlung der langfristigen Lernwirkungen vorgesehen.

## 3. Schülervorstellungen in der Sekundarstufe I

Im Rahmen einer Voruntersuchung [9] an einer Berliner Gesamtschule wurden von insgesamt 166 Schülerinnen und Schülern die Vorstellungen, die diese Schüler der Klassenstufen acht bis zehn mit dem Begriff "Hologramm" verbanden, ermittelt. Es zeigte sich, dass ca. 2/3 der Schülerinnen und Schüler keine Kenntnisse über die Bedeutung des Begriffs "Hologramm" hatten.

word hologram they linked film scenes that contained what were believed to be holograms. These alleged holographic figures, which had thus far shown properties of substantial bodies, were mainly associated with computer generation. This student body however also named as one of the primary effects the three-dimensionality of the described figures. Thus, the discussion concerning the creation of spatial images serves as a juncture for a series of lessons on holography.

Due to inexperience, none of the students naturally connected holography with interference optics.

### 4. Student conceptions in secondary school level II
Whereas at the intermediary level large areas of optics can neither be explained with the available resources nor with models, these optical phenomenon of the secondary school level I have to therefore be collected, categorized and described [5], the demands increase as students transfer into the secondary school level II.

The students are expected to explain optic events with the help of models that are partially contradictory. The transfer form the ray model to the wave model provides approaches for numerous and often unintended student concepts that only superficially reveal physical reality. Hence three examples are introduced in the following in order to show how this problem reveals itself during a lesson on holography.

Here we are firstly dealing with the discussion on three-dimensionality that took place in the portrayed series of lessons [10] in an experiential and student-oriented manner. The ray model was used to explain the observed effects.

The main conceptions of light found during student interviews, which will be analysed in the following section, are the foundation for a understanding of the problems in dealing with holography also discernible from the subsequent third example. These were worked out in the lesson based on the wave model.

### 5.a Analysis of student conceptions of space
An evaluation of the test questions indicated that even honour students have difficulties in correctly describing the production of an image on a concave mirror.

When drawn, the image is often positioned on the surface of the concave mirror. Such a naïve explanation not only reduces a three-dimensional mirror surface to one that is two-dimensional, preventing a correct comprehension of space, but also produces conceptually significant consequences. According to the perception of these students, a point of an image does not result from light focussing on one point, but from a process of projection that is intuitively ampli-

Lediglich etwas mehr als zehn Prozent der Befragten gab an, bereits einmal ein reales Hologramm gesehen zu haben. Diese Schülerinnen und Schüler konnten die wesentlichen Eigenschaften der Dreidimensionalität holographischer Effekte im Rahmen ihrer fachsprachlich noch lückenhaften Ausdrucksweise angeben.

Weit ausführlichere Beschreibungen lieferte jedoch die dritte Teilpopulation, die aus mehr als zwanzig Prozent der Schülerinnen und Schüler bestand. Sie verknüpften mit dem Begriff des Hologramms filmische Szenen, in denen vermeintliche Hologramme dargestellt wurden. Zumeist wurden diese angeblich holographischen Figuren, die bisweilen Eigenschaften materiebehafteter Körper zeigten, mit der Generierung per Computer in Verbindung gebracht. Aber auch diese Schülerpopulation nannte als eine der wesentlichen Effekte die Dreidimensionalität der beschriebenen Figuren. Als didaktischer Anknüpfungspunkt bietet sich deshalb die Diskussion der Entstehung räumlicher Bilder bei einer Unterrichtsreihe zur Holographie an.

Aufgrund fehlender Kenntnisse brachte naturgemäß keiner der Schüler die Holographie mit interferenzoptischen Effekten in Zusammenhang.

### 4. Schülervorstellungen in der Sekundarstufe II
Während in der Mittelstufe große Bereiche der Optik nicht mit den zur Verfügung stehenden Mitteln und Modellen ge- und erklärt werden können und deshalb optische Phänomene im Physikunterricht der Sekundarstufe I hauptsächlich gesammelt, geordnet und beschrieben werden [5], findet beim Übergang in die Sekundarstufe II ein Anforderungssprung statt.

Nunmehr wird von den Lernenden erwartet, optische Erscheinungen mit Hilfe von - sich teilweise widersprechenden - Modellen zu erklären. Insbesondere bietet der Übergang vom Strahlenmodell zum Wellenmodell Ansatzpunkte für zahlreiche, oft unbewusste Schülerkonzepte, die die physikalische Realität oft nur vordergründig erfassen. Deshalb wird im folgenden exemplarisch an drei Beispielen vorgestellt, wie sich diese Problematik während des Unterrichts zur Holographie zeigt.

Dabei handelt es sich zum ersten um die Diskussion der Dreidimensionalität, die in der beschriebenen Unterrichtsreihe [10] experimentell und schülerorientiert erfolgte. Die Erklärung der beobachteten Effekte wurde dabei im Strahlenmodell vorgenommen.

Die in den Schülerinterviews aufgefundenen wesentlichen Vorstellungen zum Licht, die im darauf folgenden Abschnitt analysiert werden, sind Grundlage für ein Verständnis der im sich anschließenden dritten Beispiel sichtbaren Probleme bei der Behand-

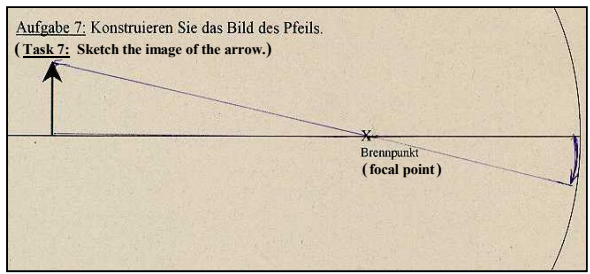

Aufgabe 7: Konstruieren Sie das Bild des Pfeils.
( **Task 7:** Sketch the image of the arrow.)

X
Brennpunkt
( focal point )

***Figure 3:*** *Student sketch showing the image formation on the concave mirror.*

fied by the inadvertently necessary absorption of the impacting light (a light ray ending on the surface of the mirror).

This hypothesis of projection that numerous students have can prevent them from being able to distinguish between real and virtual images as the function of the surface of the concave mirror is associated with that of a screen. These students ultimately relate to the production of an image which corresponds with the representation in [2]: *"Their remarks indicated, that they could not conceive of an image as existing in free space, independent of a surface."*

And since our lessons on holography rely on explaining optical illusions on carved figures in a didactically reconstructed manner by using the behaviour of light on small cylindrical concave mirrors, such conceptions have to be dealt with in class discussions. Only then can an expanded situation with a diagonal incidence of light be clarified by students in a technically accurate manner and the spatial location of a image point successfully discussed.

**5b. Analysis of student conceptions of light**
In the interviews following the series of lessons the students were asked explicitly about their conceptions of light. They regularly referred to the common model conceptions such as the ray model, the wave model and the particle model. However, the way this disposed knowledge is interpreted depends on the students and their competencies. As a result, students with meta-conceptional capabilities are able to switch from one level and method of argumentation of one model to another consciously and problem-oriented. They are not only able to designate these changes, but are also able to substantiate their claim.

Students without this level of expertise also switch between models, yet are not able to explicitly recognize these switches and demonstrate a certain unreflected arbitrariness in their methods of argumentation. Typical for this is that a phenomenon, which had been explained using a model, is then explained using another model of light. These types lung der Holographie. Diese wurde im Unterricht auf der Basis des Wellenmodells erarbeitet.

**5.a Analyse von Schülervorstellungen zur Räumlichkeit**
Wie die Auswertung der Testfragen zeigte, haben selbst Leistungskursschüler Probleme mit der korrekten Beschreibung der Bildentstehung am Hohlspiegel.

Nicht selten wird der Ort des Bildes zeichnerisch an der Oberfläche des Hohlspiegels lokalisiert. Solche naiven Erklärungsmuster reduzieren eine dreidimensionale Situation nicht nur auf die zweidimensionale Spiegeloberfläche und verhindern eine korrekte Erfassung der Räumlichkeit.

Sie haben auch konzeptuell wesentliche Folgen: Ein Bildpunkt entsteht nach Auffassung dieser Schüler nicht durch eine Fokussierung von Licht in einem Punkt, sondern durch einen Projektionsprozess, der intuitiv noch durch die unbewusst vorausgesetzte Absorption des auftreffenden Lichts (an der Spiegeloberfläche endender Lichtstrahl) verstärkt wird.

Diese Projektionshypothese etlicher Schülerinnen und Schüler kann eine Nichtunterscheidbarkeit reeller und virtueller Bilder induzieren, denn der Oberfläche des Hohlspiegels wird dadurch die Funktion eines Schirms zugeordnet. Letztendlich haben diese Schüler eine Vorstellung bezüglich der Bildentstehung, die sich mit den Beschreibungen in [2] decken: *„Ihre Bemerkungen deuten darauf hin, dass sie sich ein Bild im freien Raum schwebend, unabhängig von einem Schirm bzw. einer Oberfläche, nicht vorstellen können."*

Da in der Unterrichtseinheit zur Holographie die didaktisch reduzierte Erklärung der optischen Erscheinungen an Ritzfiguren durch Rückgriff auf das Verhalten von Licht an schmalen zylindrischen Hohlspiegeln erfolgt, müssen solche Vorstellungen im Unterrichtsgespräch aufgearbeitet werden. Erst dann kann die auf einen schrägen Lichteinfall erweiterte Situation durch Schüler fachlich korrekt geklärt und die räumliche Lage eines Bildpunktes erfolgreich diskutiert werden.

**5.b Analyse von Schülervorstellungen zum Licht**
In den Interviews nach Abschluss der Unterrichtsreihe wurden Schüler explizit nach ihren Vorstellungen zum Licht befragt. Regelmäßig nannten sie dabei die gängigen Modellvorstellungen wie Strahlenmodell, Wellenmodell und Teilchenmodell. Die Interpretation dieses Verfügungswissens erfolgt jedoch durch Schüler unterschiedlicher Leistungsstärke in gänzlich unterschiedlicher Art und Weise. So sehen sich Schüler mit metakonzeptionellen Fähigkeiten in der Lage, bewusst und problemorientiert von Argumentationsmustern eines Modells in die Ebene eines anderen Modells zu springen und of students have the tendency to link the characteristics of both models in such a way that the created hybrid model seems all encompassing.

Using the example of convergence on the concave mirror, which was explained by students via the ray model without trouble both verbally and by drawing, figures 4 a through c exemplify this undertaking. The student demonstrates a classic school book situation in image 3 a, using the ray model. In image 3 b, the student does not seem willing to abandon the primary properties of the ray model (an obvious light path) as he tries to make the light rays fit the wave model of light using a wavy structure.

After being asked to possibly explain using the particle model, the student eventually modifies his drawing in such a way (figure 4 c) so that the light particles follow along the sinus-shaped light path and are steered into a focus. This non-existent distinction between the discussed light models hinders an adequate utilization of models.

diese Sprünge nicht nur zu benennen, sondern auch zu begründen.

Schüler ohne dieses Expertenniveau springen gleichfalls zwischen Modellen hin und her, erkennen jedoch diese Sprünge nicht explizit und weisen in ihren Argumentationsmustern eine unhinterfragte Beliebigkeit auf. Typisch ist dabei das Vorgehen, wenn ein Phänomen, das mit Hilfe eines Modells erläutert wurde, nun mit Hilfe eines anderen Lichtmodells erklärt werden soll. Dabei zeigt sich bei diesen Schülern die Tendenz, Charakteristika beider Modelle so zu verknüpfen, dass das entstehende Hybridmodell allumfassend scheint.

Am Beispiel der Brennpunktbildung am Hohlspiegel, die durch die Schüler im Strahlenmodell problemlos sowohl zeichnerisch wie auch verbal erläutert werden konnte, zeigen die Bilder 4 a bis c dieses Vorgehen exemplarisch. In Bild 3 a stellt der Schüler die klassische Lehrbuchsituation dar, wobei er sich auf das Strahlenmodell bezieht. In Bild 3 b ver-

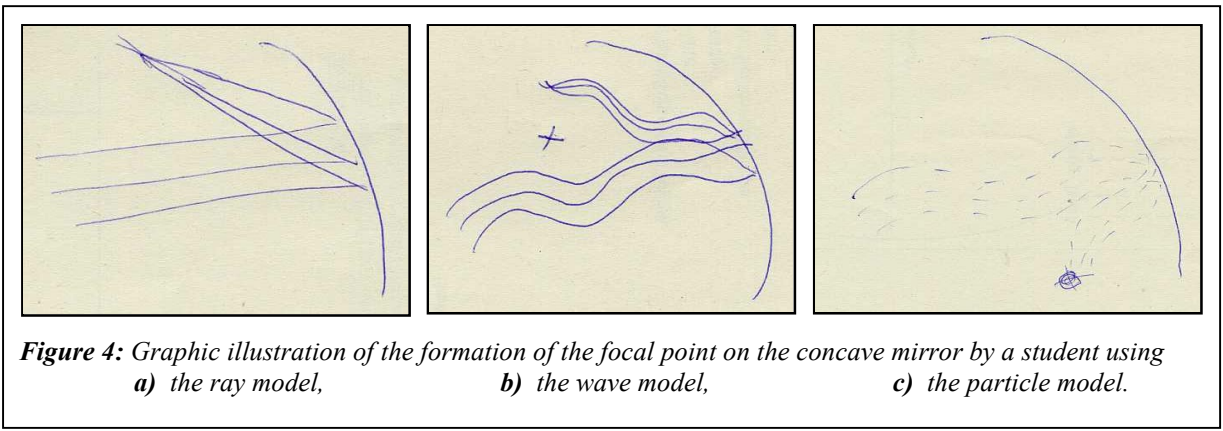

**Figure 4:** *Graphic illustration of the formation of the focal point on the concave mirror by a student using*
  ***a)*** *the ray model,*          ***b)*** *the wave model,*          ***c)*** *the particle model.*

### 5.c Analysis of student conceptions on interference optics

Numerous scenarios in class brought about implicit and explicit conceptions of students concerning interference. Pre-concepts, as illustrated by Ambrose et al. [1] in their very extensive study, were partially confirmed.

To a great extent inaccuracies have their origins in a terminological jumble of not jet established expressions. Consequently, uncertainties frequently surfaced in the students' arguments if they did not correctly separate the meaning of technical terms. The overview in figure 5 gives several repeatedly observed examples.

| | | |
|---|---|---|
| **wave** | **vs.** | **wave front** |
| **coherence length** | **vs.** | **wave length** |
| **elementary wave** | **vs.** | **object wave** |

**Figure 5:** *Terminology students had problems separating based on their content.*

sucht er, die Lichtstrahlen durch eine wellenförmige Struktur dem Wellenmodell des Lichts anzupassen, ohne eine der wesentlichsten Eigenschaften des Strahlenmodells (eineindeutiger Lichtweg) aufgeben zu wollen.

Letztendlich modifiziert der Schüler auf Nachfrage nach einer eventuellen Erklärung im Teilchenmodell seine Zeichnung so (Bild 4c), dass die Lichtteilchen den sinusförmigen Lichtwegen entlang laufen und in einen Brennpunkt gelenkt werden. Diese nichtvorhandene Trennschärfe zwischen den diskutierten Lichtmodellen verhindert einen adäquaten Modellgebrauch.

### 5.c Analyse von Schülervorstellungen zu Interferenzeffekten

Zahlreiche Unterrichtsszenen brachten implizite und explizite Vorstellungen von Schülern zu Interferenzerscheinungen zu Tage. Teilweise wurden Präkonzepte, wie sie Ambrose et al. [1] in ihrer sehr umfangreichen Studie darlegten, bestätigt.

As the following example shows, students had troubles concerning abstract representations and graphic interpretations.

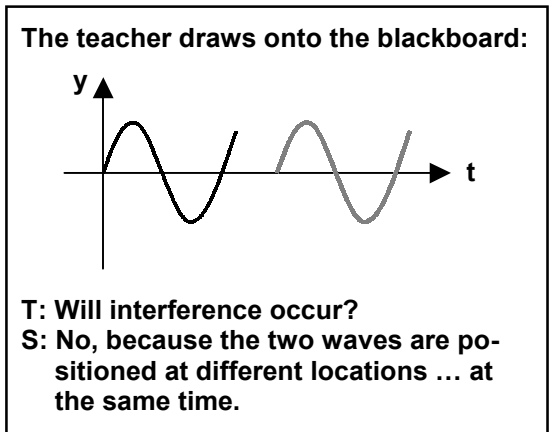

> **The teacher draws onto the blackboard:**
>
> **T: Will interference occur?**
> **S: No, because the two waves are positioned at different locations … at the same time.**

**Figure 6:** In-class scenario concerning coherence.

Statements of this type from students not only confirm deficits in the mathematic-analytical sphere, but also point to a didactically deep seated problem. Many students seem to have difficulties in principally separating the spatial and temporal observations when using the wave theory. This random switching of space and time, which manifests itself in student's statements ("It doesn't matter what is applied to the x-axis.") as well as in the inability to distinguish between oscillations and waves, leads to further problems concerning holography as standing waves are discussed.

In addition, it became evident during lessons on holography that the switch from one-dimensional waves, which had been established in the previous lesson on interference optics, caused problematic perspectives concerning spatial wave fields. Consequently, some students questioned whether diagonally colliding coherent light can produce in-

Ein großer Bereich von Fehlleistungen hat ihre Ursache in einer Begriffsmischung noch nicht gefestigter Ausdrücke. So traten häufiger Unklarheiten in der Argumentation von Schülerinnen und Schülern auf, wenn diese Fachbegriffe in der Bedeutung nicht korrekt trennen. Einige oft beobachtete Beispiele zeigt die folgende Übersicht:

> **Welle         ./. Wellenfront**
> **Kohärenzlänge ./. Wellenlänge**
> **Elementarwelle ./. Objektwelle**

Die inhaltliche Trennung dieser Begriffspaare misslang den Schülerinnen und Schülern oft. Ebenfalls nachweisbar sind Überforderungen der Schüler im Bereich abstrakter Darstellungen und graphischer Interpretationen, wie folgendes Beispiel, das sich auf Abbildung 6 bezieht, zeigt:

> **L: Kommt es zur Interferenz?**
> **S: Nein, weil die zwei Wellen örtlich unterschiedlich gelagert sind ... zum selben Zeitpunkt.**

Schüleräußerungen dieser Art belegen jedoch nicht nur Defizite im mathematisch-analytischen Bereich, sondern weisen auch auf ein didaktisch tiefer gehenderes Problem hin. So scheint für viele Schüler im Bereich der Wellentheorie kein prinzipieller Unterschied zwischen räumlichen und zeitlichen Betrachtungen zu bestehen. Diese beliebige Austauschbarkeit von Raum und Zeit, die sich sowohl in Schüleräußerungen ("Es ist doch egal, was man auf der x-Achse aufträgt.") wie auch in der Ununterscheidbarkeit von Schwingungen und Wellen durch manche Schüler manifestiert, führt zu weiteren Problemen im Bereich der Holographie, wenn stehende Wellen diskutiert werden.

> **Student 1: The wave field is stored by illuminating the object using the light of a standing wave.**
>
> **Student 2: So reference wave and object wave have to both be standing waves *(short pause)* … or does that only happen when both waves are superimposed?**

> **Schüler 1: Das Wellenfeld speichert man, indem man das Objekt mit Licht einer stehenden Welle beleuchtet.**
>
> **Schüler 2: Also müssen Referenzwelle und Objektwelle beides stehende Wellen sein *(kurze Pause)* ... oder entsteht das erst bei Überlagerung der beiden Wellen?**

**Figure 7:** *Student statements during a discussion how to record a hologram.*

terference effects. As they were constructing their arguments, they were also clearly stuck in patterns of thought, which assumed that light must be parallel.

Due to the restriction to one-dimensional observations, the typical conceptions of some students were

Darüber hinaus war im Unterricht zur Holographie festzustellen, dass der Übergang von eindimensionalen Wellen, die durch den vorangegangenen Unterricht zur Interferenzoptik gefestigt vorlagen, auf räumliche Wellenfelder problematische Anschauungen auslöste. So wurde von manchen Schülern be-that the interference conditions had to be fulfilled for each distance. In response to corresponding inquiries, students often did not recognize that the difference of the optical path length are even or odd multiples of the wave length. The equalization of a retardation with an optical path length leads to the fact that this optical path length is understood as a type of resonator.

## 6. Consequences for physics education

The results introduced in the previous sections show that the promotion of meta-conceptional utilization of models is one of the most urgent concerns of present day physics lessons. For this is not only necessary for students to be able to switch models for different problems in class. To boot, experiences of the students of the limits of models should not be suppressed.

For holography this means that the separation between the ray model and the wave model should take place in a clear and structured manner. Lesson plans that suggest a combination of both levels of models (e.g. it is suggested in [6]: *"By implementing the laws and perspectives of geometric optics on zone plates it is possible to arrive at quantitative assertions."*) carry the great danger that this combination of models is accepted by the students without questions.

In conclusion, the following results can be formulated for lessons of physics:

⟹ Meticulous phases of clarification are necessary during lessons, in which students are able to present their conceptions without scrutiny by the teacher. The concealment and suppression of previous concepts and naïve student conceptions leads to an ulterior consolidation of these concepts.

⟹ Working through model conceptions by way of deliberate discussion and solution of physical questions with the help of alternative approaches, especially on different levels and paths of models as well as the working into meta-conceptional approaches.

⟹ Application of computer simulation programs that allow a model switch during the simulation process.

⟹ Integration of switching models into the everyday experiences of the students.

zweifelt, dass schräg aufeinandertreffendes kohärentes Licht Interferenzeffekte zeigt. Auch waren sie in ihren Argumentationsmustern stark in Denkmustern verhaftet, die eine Parallelität des Lichts zur Voraussetzung hatte.

In der Folge ergab sich durch die Beschränkung auf eindimensionale Betrachtungsmuster die typische Vorstellung einiger Schülerinnen und Schüler, dass die Interferenzbedingungen jeweils für eine Wegstrecke zu erfüllen seien. Auf entsprechende Nachfragen wurde von Schülern oft nicht erkannt, dass Weglängendifferenzen geradzahlige oder ungradzahlige Vielfache der Wellenlänge sein sollen. Diese Gleichsetzung eines Gangunterschieds mit einem Lichtweg führt dazu, dass dieser Lichtweg dann als eine Art Resonator aufgefasst wird.

## 6. Didaktische Konsequenzen
Die im den vorangegangenen Abschnitte vorgestellten Ergebnisse zeigen, dass eine Förderung metakonzeptionellen Modellgebrauchs eine der vordringlichsten Anliegen eines zeitgemäßen Physikunterrichts sein sollte. Dabei sind im Unterricht nicht nur Modellwechsel bei unterschiedlichen Problemstellungen durch die Schüler zu erarbeiten. Darüber hinaus darf eine Erfahrung von Modellgrenzen durch die Schülerinnen und Schüler nicht unterschlagen werden.

Im Bereich der Holographie heißt dies, dass die Abgrenzung zwischen Strahlenmodell und Wellenmodell strukturiert und klar erfolgen sollte. Unterrichtsvorschläge, die eine Vermischung beider Modellebenen fordern (z.B. wird in [6] vorgeschlagen: „*Die Anwendung der Gesetze und Betrachtungsweisen der geometrischen Optik auf die Zonenplatten macht es möglich, zu quantitativen Aussagen zu kommen.*") bergen die große Gefahr, dass diese Modellmischung unkritisch durch Schüler übernommen wird.

Zusammenfassend lassen sich die folgenden Konsequenzen für den Physikunterricht formulieren:

⟹ Notwendigkeit von ausführlichen Erläuterungsphasen im Unterricht, in denen Schüler ohne Lehrerbewertung ihre Vorstellungen vortragen können und sollen. Verschweigen und Unterdrücken von Vorkonzepten und naiven Schülervorstellungen bedeutet eine untergründige Festigung dieser Konzepte.

⟹ Aufarbeitung von Modellvorstellungen durch bewusste Diskussion und Problemlösung physikalischer Fragestellungen mit Hilfe alternativer Zugänge und insbesondere auf verschiedenen Modellebenen und verschiedenen Modellwegen sowie die Notwendigkeit der Einübung in meta-konzeptionelle Denkweisen.

⇒ Einsatz von Computer-Simulationsprogrammen, die einen Modellwechsel bei der Darstellung des Simulationsprozesses zulassen.

⇒ Einbindung von Modellwechseln in die lebensweltlichen Erfahrungen der Schülerinnen und Schüler.

## 7. Literaturangaben

The German version is published at: